%% file: spacecomp.tex
\documentclass[conference]{IEEEtran}
\IEEEoverridecommandlockouts
\usepackage{cite}
\usepackage{amsmath,amssymb,amsfonts}
\usepackage{algorithmic}
\usepackage{graphicx}
\usepackage{textcomp}
\usepackage{xcolor}
\usepackage{balance}
\usepackage{booktabs}
\usepackage{eurosym}
\usepackage{enumerate}

\newcommand{\para}[1]{\vspace{1.5mm}\noindent\textbf{#1}}

\def\BibTeX{{\rm B\kern-.05em{\sc i\kern-.025em b}\kern-.08em
    T\kern-.1667em\lower.7ex\hbox{E}\kern-.125emX}}
\begin{document}

\title{Lightspeed Data Compute for the Space Era}

\author{\IEEEauthorblockN{}
\IEEEauthorblockA{\textit{Computer Science Department} \\
\textit{RISE - Research Institutes of Sweden}\\
Stockholm, Sweden}}

\author{\IEEEauthorblockN{
    Thomas Sandholm\IEEEauthorrefmark{1},
    Bernardo A. Huberman\IEEEauthorrefmark{1}, Klas Segeljakt\IEEEauthorrefmark{1}, and
    Paris Carbone\IEEEauthorrefmark{1}\IEEEauthorrefmark{2}
    }
    
\IEEEauthorblockA{\IEEEauthorrefmark{1}
\textit{RISE - Research Institutes of Sweden}\\
    thomas.sandholm@ri.se, bernardo.huberman@ri.se, klas.segeljakt@ri.se, paris.carbone@ri.se
    }
\IEEEauthorblockA{\IEEEauthorrefmark{2}
  \textit{KTH Royal Institute of Technology}\\
    parisc@kth.se
    }
}

\maketitle

\input{abstract}
\input{introduction}
\input{background}
\input{overview}
\input{model}

\input{simulation}
\input{conclusion}

\balance
\bibliographystyle{plain}
\bibliography{related}

\end{document}

%% file: abstract.tex
\begin{abstract}
While thousands of satellites photograph Earth every day, most of that data never makes it to the ground because downlink bandwidth simply cannot keep up. Processing data in the Low Earth Orbit (LEO) zone offers promising capabilities to overcome this limitation.  We propose \emph{SpaceCoMP}, a MapReduce-inspired processing model for LEO satellite mesh networks. Ground stations submit queries over an area of interest; satellites collect sensor data, process it cooperatively at light-speed using inter-satellite laser links, and return only the results. Our compute model leverages space physics to accelerate computations on LEO megaconstellations. Our distance-aware routing protocol exploits orbital geometry. In addition, our bipartite match scheduling strategy places map and reduce tasks within orbital regions while minimizing aggregation costs. We have simulated constellations of 1,000--10,000 satellites showcasing 61--79\% improvement in map placement efficiency over baselines, 18--28\% over greedy allocation, and 67--72\% reduction in aggregation cost. SpaceCoMP demonstrates that the orbital mesh is not merely useful as a communication relay, as seen today, but can provide the foundations for faster data processing above the skies.
\end{abstract}


%% file: introduction.tex
\section{Introduction}

Low-earth orbit (LEO) satellite constellations are generating data at rates that far exceed their capacity to transmit to the ground. A single high-resolution Earth observation satellite produces 1--2~TB of imagery per day~\cite{selva2012cubesats}, yet typical ground station contacts, lasting only 5--15 minutes per orbit, can transfer merely a fraction of this volume over bandwidth-constrained radio-frequency (RF) downlinks. With mega-constellations now reaching unprecedented scale, Starlink exceeded 7,500 operational satellites in 2024 with over 40,000 planned, and the EU's IRIS\textsuperscript{2} program commits \euro10.6B to European sovereign connectivity~\cite{mcdowell2024starlink,iris2024}, the aggregate data generation potential reaches petabytes per day. Yet global ground station infrastructure remains the bottleneck: RF spectrum is shared and regulated, weather causes signal attenuation, and ground stations must be within line-of-sight during brief orbital passes. This \emph{downlink wall} fundamentally limits the value that can be extracted from space-based sensing.

The emergence of optical inter-satellite links (ISL) offers a transformative opportunity. Unlike RF links that must traverse Earth's atmosphere, optical ISLs operate in vacuum at the speed of light, approximately 47\% faster than signals in terrestrial fiber~\cite{handley2018delay}, with bandwidth exceeding 10~Gbps per link~\cite{perdigues2021hydron}. Table~\ref{tab:rf-vs-isl} summarizes the key differences. Modern LEO satellites equipped with laser terminals form mesh networks where any satellite can route data to any other through a sequence of hops, creating a distributed infrastructure with high-bandwidth internal connectivity but constrained external bandwidth to ground. From a computational lense the LEO architecture features thousands of interconnected compute nodes and data is naturally partitioned across collectors. The cost of performing external I/O essentially mirrors similar conditions that motivated MapReduce in terrestrial data centers two decades ago~\cite{dean2008}. A key insight is that \emph{processing data in orbit before transmission can dramatically reduce the potential of downlink saturation}, converting terabytes of raw sensor data into megabytes of actionable results.

\begin{table}[t]
\centering
\caption{RF Ground Links vs.\ Optical Inter-Satellite Links (ISLs)}
\label{tab:rf-vs-isl}
\renewcommand{\arraystretch}{1.5}
\setlength{\tabcolsep}{8pt}  
\begin{tabular}{@{}lll@{}}
\toprule
\textbf{Property} & \textbf{RF Ground Links} & \textcolor{violet}{\textbf{Optical ISLs}} \\
\midrule
Medium & Atmosphere (lossy) & Vacuum \\
\addlinespace[0.3em]
Propagation speed & $\approx$200,000 km/s & 299,792 km/s (\textcolor{violet}{$c$}) \\
\addlinespace[0.3em]
Bandwidth & Shared spectrum, $<$1 Gbps & 10--100+ Gbps/link \\
\addlinespace[0.3em]
Availability & Weather-dependent & Always-on \\
\addlinespace[0.3em]
Ground infrastructure & Required & Not required \\
\addlinespace[0.3em]
Long-haul latency & High (via ground hops) & Low (direct path) \\
\addlinespace[0.3em]
Atmospheric loss & Yes (rain, clouds) & None \\
\bottomrule
\end{tabular}
\end{table}

However, distributed computing in LEO presents novel challenges absent from terrestrial systems. Satellites orbit at 7.5~km/s, completing a full orbit in approximately 95 minutes. The network topology changes continuously: inter-plane link distances vary by up to 40\% over each orbital period as planes converge near the poles and diverge at the equator. Near polar regions, satellites from adjacent planes pass each other at relative velocities exceeding 15~km/s, forcing temporary link disconnection. Task placement must therefore account not just for hop count but for actual transmission distance and time-varying link geometry. Furthermore, satellites remain in proximity to a ground station for only 5--15 minutes before handover~\cite{pfandzelter2023leo}, requiring coordination of data collection, processing, and result delivery within narrow windows. Existing cloud software, designed for static datacenter topologies, cannot be directly deployed: the assumptions underlying conventional schedulers, routers, and consistency protocols break down when the infrastructure itself is in constant motion.

We propose Space Collect-MapReduce Processing (SpaceCoMP), a distributed processing model designed for LEO mega-constellation ISL networks. Consider hundreds of satellites capturing imagery of a wildfire. How do we extract intelligence from such a rich, distributed dataset without saturating the downlink? The data is already naturally partitioned across collector satellites; the challenge is ensuring that computation happens close to the data with minimal shuffling, both within the orbital network and ultimately when transmitting results to the ground. SpaceCoMP extends the MapReduce paradigm with a \emph{Collect} phase that explicitly models sensor data acquisition from an area of interest (AOI), followed by Map processing on nearby satellites and Reduce aggregation before consuming downlink bandwidth.

Our approach draws inspiration from MapReduce~\cite{dean2008}, the NASA core Flight System (cFS)\footnote{https://etd.gsfc.nasa.gov/capabilities/core-flight-system/}, and Wireless Sensor Network protocols such as Contract-Net~\cite{smith1980} and Directed Diffusion~\cite{intanagonwiwat2003}. We make the following contributions:

\begin{enumerate}
    \item \textbf{The SpaceCoMP processing model}: A three-phase (Collect-Map-Reduce) abstraction for data-intensive workloads in LEO constellations that lets developers focus on application logic while the infrastructure handles orbital dynamics and task placement (Section~\ref{sec:overview}).
    
    \item \textbf{Distance-optimized routing}: A routing algorithm that exploits time-varying cross-plane link geometry to minimize total transmission distance by 8--21\% (depending on orbital inclination) without increasing hop count (Section~\ref{sec:routing-eval}).
    
    \item \textbf{Bipartite matching scheduler}: A map task allocation strategy based on maximum-weight matching in bipartite graphs, improving placement efficiency by 61--79\% over random allocation and 18--28\% over greedy baselines (Section~\ref{sec:map-eval}).
    
    \item \textbf{Reduce placement strategy}: A center-of-AOI reducer placement that reduces aggregation cost by 67--72\% compared to routing results directly to the line-of-sight ground station (Section~\ref{sec:e2e-eval}).
\end{enumerate}

We validate SpaceCoMP through simulations modeling Walker Delta constellations with realistic ISL parameters across constellation sizes from 1,000 to 10,000 satellites.

%% file: background.tex
\section{From Cloud to Space Computing: a Primer}

This section introduces the LEO environment, including a primer of satellite networking (Section~\ref{sec:leo-primer}). We then position our work relative to terrestrial cloud computing~(\ref{sec:cloud-primer}) and sensor and edge network~(\ref{sec:edge-primer}) literature.

\subsection{The Low-Earth-Orbit Computing Environment}
\label{sec:leo-primer}

Satellite mega-constellations share similarities with datacenter networking: fat-tree topologies, rack-aware placement, millisecond latencies, and multi-Gbps bandwidths. In particular, LEO satellite networks share the goal of interconnecting compute nodes but operate under fundamentally different physical constraints and capabilities possible in vacuum.

\subsubsection{Orbital Altitude and Its Implications}

Earth-orbiting satellites occupy three primary altitude bands. Geostationary Earth Orbit (GEO) satellites at 35,786~km remain fixed relative to ground locations but suffer approximately 120~ms one-way latency. Medium Earth Orbit (MEO), used by GPS, spans 8,000--20,000~km with orbital periods of 2--12 hours. LEO, at altitudes of 300--2,000~km, has emerged as the preferred domain for mega-constellations due to properties critical for distributed computing~\cite{prol2022leo,yin2025orbital}:

\para{Low latency}: At typical altitudes of 500--600~km, one-way propagation delay to ground is only 2--4~ms, enabling round-trip times of approximately 10~ms~\cite{izhikevich2024democratizing}, comparable to inter-datacenter latencies.
    
\para{Short orbital period}: LEO satellites complete an orbit in 90--95 minutes, traveling at 7.5~km/s and circling the globe roughly 15 times per day.
    
\para{Distributed coverage}: Each satellite's ground footprint spans only $\sim$1,000~km diameter, necessitating constellations of hundreds to thousands of satellites while also enabling data-parallel data access by nature across nodes.

\subsubsection{Constellation Geometry}

LEO satellites are organized into structured formations that determine network topology and thus task placement options:

\para{Orbital planes and shells.} Satellites following the same circular path form an \emph{orbital plane}, maintaining fixed spacing relative to one another. Multiple planes sharing the same altitude and inclination, distributed uniformly around Earth, constitute a \emph{shell}. The Walker Delta pattern~\cite{walker1984satellite} arranges $N$ planes separated by $360^{\circ}/N$ in longitude, producing a ``doughnut-shaped'' coverage volume with wraparound in both axes, naturally forming the 2D-Torus topology that can be leveraged for task placement.

\para{Inclination.} The angle between an orbital plane and Earth's equator determines coverage latitude. For instance, Starlink uses 53° inclination for populated regions; Earth observation missions often use near-polar orbits (e.g., 87° as in our simulations) for global coverage.

\para{Phase offset.} Satellites across planes are staggered to maximize coverage uniformity. This geometric regularity enables SpaceCoMP's bipartite matching formulation: when an area of interest maps to a rectangular region of satellites, tasks and processors form a natural grid structure.

\subsubsection{Inter-Satellite Links}

The emergence of optical inter-satellite links (ISLs) has transformed LEO constellations from isolated satellites into interconnected mesh networks~\cite{perdigues2021hydron}. Operating in vacuum via laser terminals, ISLs achieve:

\begin{itemize}
    \item \textbf{Speed}: Light travels at \textcolor{violet}{$c$} $=299{,}792$~km/s in vacuum, approximately 47\% faster than in optical fiber~\cite{handley2018delay}.
    \item \textbf{Bandwidth}: Modern terminals achieve 10--100+ Gbps per link.
    \item \textbf{Availability}: Continuous operation independent of weather or ground station visibility.
\end{itemize}

\textbf{The +Grid topology.} LEO satellites typically carry four laser terminals connecting to neighbors ahead and behind in the same plane, plus neighbors in adjacent planes left and right. This forms a +Grid (2D-Torus) topology~\cite{pfandzelter2022qos} where routing follows Manhattan distances, i.e., the number of hops between satellites equals the sum of horizontal (cross-plane) and vertical (within-plane) steps.

\subsubsection{Dynamic Topology}

Unlike datacenter networks, LEO mesh topology changes continuously but \emph{predictably}:

\textbf{Static intra-plane distances.} Satellites within a plane maintain formation; inter-satellite distance is constant (Equation~\ref{eq:intra-plane}).

\textbf{Cyclic inter-plane distances.} Orbital planes converge near poles and diverge at the equator. Cross-plane link distances vary sinusoidally over each orbital period (Equation~\ref{eq:inter-plane}), exceeding 40\% variation for high-inclination orbits.

\textbf{Polar crossover events.} Near poles, satellites from adjacent planes pass each other at relative velocities exceeding 15~km/s. Cross-plane links must be disabled during these ``seam'' crossings~\cite{pfandzelter2023leo}.

\textbf{Ascending vs.\ descending trajectories.} Half the satellites in a plane travel northward (ascending), half southward (descending). Their extreme relative velocity prevents stable ISL links between them. This reveals an implicit mutual exlusion constraint when schedulling compute: it is only possible to select \emph{only} ascending or \emph{only} descending satellites for any computation, but not a mix.

Critically, all dynamics are deterministic given orbital parameters. Satellite positions can be predicted in advance using propagation models such as SGP4~\cite{vallado2006revisiting}. It is therefore crucial to exploit this orbital predictability in both routing and task placement when performing computing in space.

\subsection{Cloud Computing Background}
\label{sec:cloud-primer}

SpaceCoMP draws on three research threads in terrestrial cloud computing: distributed task scheduling, the MapReduce programming model, and wireless sensor networks.

\subsubsection{Distributed Task Scheduling}

The challenge of mapping tasks to processors has evolved through several eras. Early multiprocessor systems recognized that naive scheduling causes thrashing; co-scheduling techniques modeled allocation as bin-packing to avoid fragmentation~\cite{ousterhout1982}. As HPC clusters became oversubscribed, backfilling~\cite{jones1999} improved utilization by allowing small jobs to advance when they would not delay larger ones.

Cloud computing (2010--) shifted the paradigm toward horizontal scaling across commodity hardware. Kubernetes~\cite{burns2016} schedules containers based on resource requirements, affinity rules, and image locality. Apache Spark's delayed scheduling~\cite{zaharia2010} trades fairness for data locality by allowing jobs to wait briefly for preferred nodes. More recently, adaptivity has been applied to task placement for ML workloads~\cite{mirhoseini2017} e.g., through the use of reinforcement learning. This allows ``learning'' placement policies that minimize execution time in heterogeneous GPU/CPU environments.

A distinct challenge emerges: the ``datacenter'' itself is in motion. Static affinity rules are insufficient when node proximity changes continuously. Space computing schedulers should therefore capture time-varying placement costs that terrestrial schedulers otherwise treat as constant.

\subsubsection{MapReduce and Data-Parallel Processing}

In the 2010s, Google revealed MapReduce~\cite{dean2008} a compute paradigm that simplified distributed programming. While providing an over-simplified interface: \texttt{map} and \texttt{reduce} functions, MapReduce democratized automatic parallelization, fault tolerance, and data-locality-aware scheduling. Its key optimization was moving computation to data rather than vice versa, which greatly reduced costly network shuffles.

Despite the similarities exhibited in LEO satellite networks, the original MapReduce model would fail in orbit. Unlike terrestrial MapReduce, where ``local'' versus ``remote'' is binary, placement costs need to be remodelled to account for continuous functions of hop counts and physical distances. In addition, programming semantics are also missing to express pre-compute operations during sensor data acquisitions that precede processing.

\subsection{Wireless and Edge Computing}
\label{sec:edge-primer}

\subsubsection{Wireless Sensor Networks}

Earth observation via satellite constellations have been previously studied within wireless sensor networks (WSN). In this context, sattelites are edge nodes that collect and aggregate environmental data.  Problems of task allocation, relevant to this work, have been studied in similar settings. For example, the Contract-Net protocol~\cite{smith1980} introduced competitive bidding for task allocation; distributed sensors bid based on their cost to complete tasks, and a controller minimizes total cost. Van der Horst et al.~\cite{van2012} also applied similar market-based allocation to satellite networks, though not for the ISL mesh topologies now prevalent. Directed Diffusion~\cite{intanagonwiwat2003} demonstrated that routing nodes can actively aggregate results from distributed sensors, reducing data volume before transmission to a sink, in an analogous fashion to a reduce phase. In this work we capitalize on these ideas. As shown in \ref{sec:model}, it is possible to have collectors bid implicitly through the cost matrix, whereas, mappers can process locally, and reducers aggregate before the final hop to ground.

\subsubsection{LEO Edge Computing}

Recent work has begun exploring computation in LEO. Krios~\cite{bhosale2024krios} introduces ``availability zones'' that mask satellite mobility, enabling terrestrial orchestrators to schedule predictively. Celestial~\cite{pfandzelter2022celestial} provides a virtual testbed for LEO edge emulation. SkyMemory~\cite{sandholm2025} demonstrates distributed caching for transformer inference across satellites.

These systems treat LEO as an extension of terrestrial infrastructure. Yet, we aim for a programming model that embraces orbital dynamics as a first-class scheduling concern, optimizing task placement for the unique cost structure of time-varying ISL distances.

%% file: overview.tex
\section{System Overview}
\label{sec:overview}

\begin{figure}
    \centering
    \includegraphics[width=0.99\linewidth]{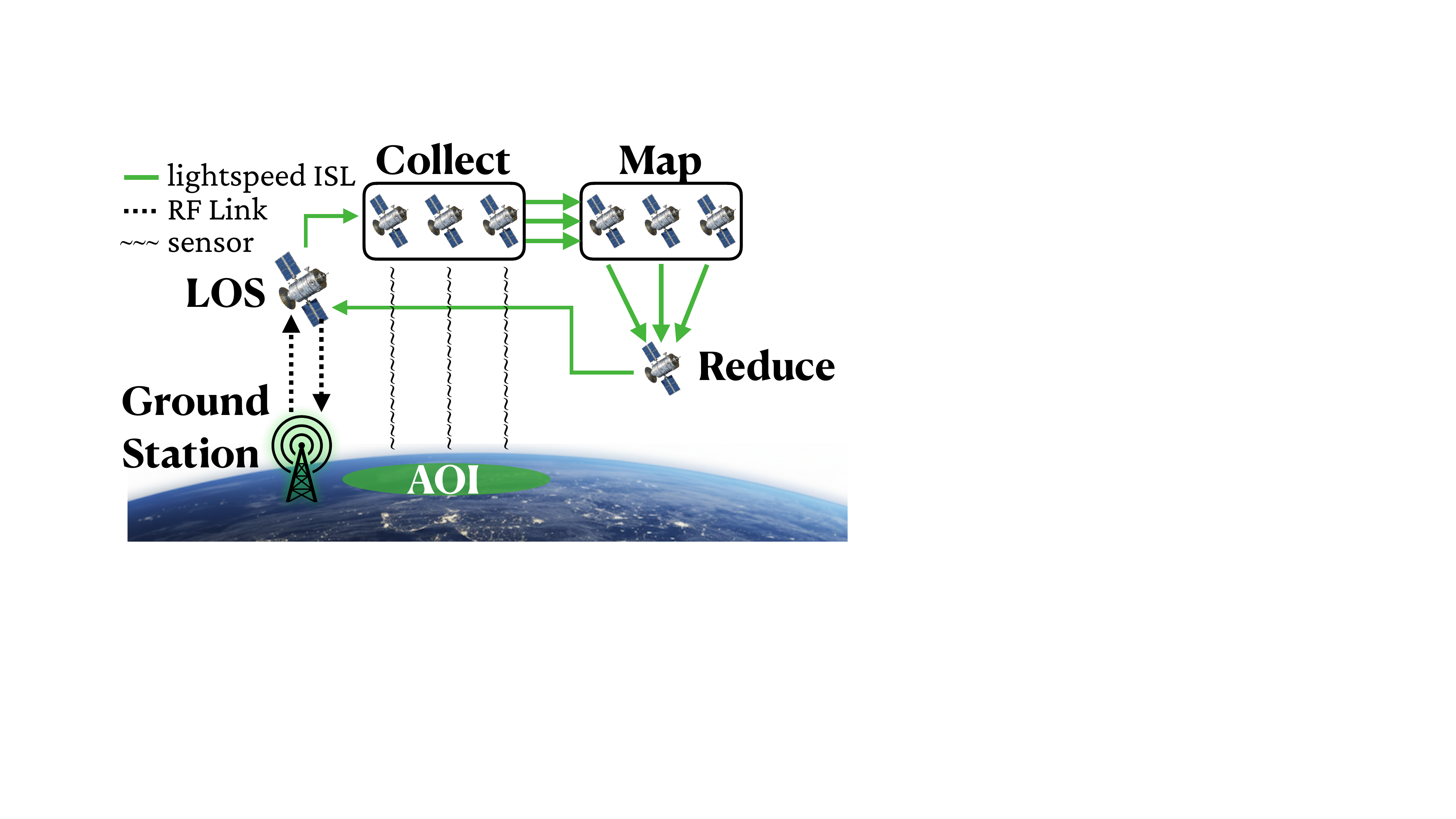}
    \caption{SpaceCoMP Architecture. Collectors observe the AOI via onboard sensors (wavy lines), exchange data with mappers and reducer via optical ISLs (solid lines), and deliver results to ground via RF links (dotted lines).}
    \label{fig:comparch}
\end{figure}

SpaceCoMP enables distributed data processing across LEO satellite constellations by extending the MapReduce paradigm with an explicit data collection phase. Figure~\ref{fig:comparch} illustrates the architecture.

\subsection{Request Flow}

A ground station initiates processing by submitting a \emph{job request} to the nearest visible satellite, termed the \emph{Line-of-Sight (LOS) node}. The request specifies:

\begin{itemize}
    \item An \textbf{Area of Interest (AOI)}: a geographic bounding box (latitude/longitude pairs) defining the region to observe.
    \item A \textbf{Collect task}: sensor configuration and data acquisition parameters.
    \item A \textbf{Map task}: local-compute logic applied to each collected data partition.
    \item A \textbf{Reduce task}: aggregation logic combining all map outputs before downlink.
\end{itemize}

The LOS node acts as the \emph{coordinator}, translating the geographic AOI into a grid of satellite nodes and orchestrating the three processing phases.

\subsection{Processing Phases}

\textbf{Phase 1: Collect.} The coordinator identifies satellites whose ground projections intersect the AOI. These \emph{collector nodes} observe the AOI using onboard sensors (e.g., optical cameras, SAR, spectrometers) and store the acquired data locally, avoiding immediate network transfer.

\textbf{Phase 2: Map.} A subset of satellites within or near the AOI are designated as \emph{mapper nodes}. The coordinator solves a bipartite matching problem (Section~\ref{sec:scheduling}) to assign each collector's data to a mapper, minimizing total transfer cost over ISL links. Mappers execute the user-defined map function on their assigned data partitions, producing intermediate results.

\textbf{Phase 3: Reduce.} A single \emph{reducer node}, selected to minimize aggregate transfer distance from all mappers (Section~\ref{sec:reduce-placement}), collects intermediate results via ISL and applies the reduce function. The final output is routed to the LOS node for RF downlink to the ground station.

\subsection{Design Rationale}

SpaceCoMP's three-phase structure addresses the unique constraints of orbital networks:

\begin{itemize}
    \item \textbf{Data locality}: By processing data near its collection point, SpaceCoMP minimizes ISL bandwidth consumption. Only reduced results traverse the long path to the LOS node.
    
    \item \textbf{Topology awareness}: Unlike terrestrial MapReduce where datacenter topology is static, SpaceCoMP's scheduler accounts for time-varying inter-plane distances when computing placement costs.
    
    \item \textbf{Downlink optimization}: The reduce phase compresses terabytes of raw observations into megabytes of actionable intelligence before the bandwidth-constrained RF downlink.
\end{itemize}

The coordinator need not be the LOS node; any satellite with sufficient visibility could orchestrate the computation. However, placing coordination at the LOS node simplifies result delivery and ground station interaction.

%% file: model.tex
\section{System Model}
\label{sec:model}

\begin{figure}
    \centering
    \includegraphics[width=0.99\linewidth]{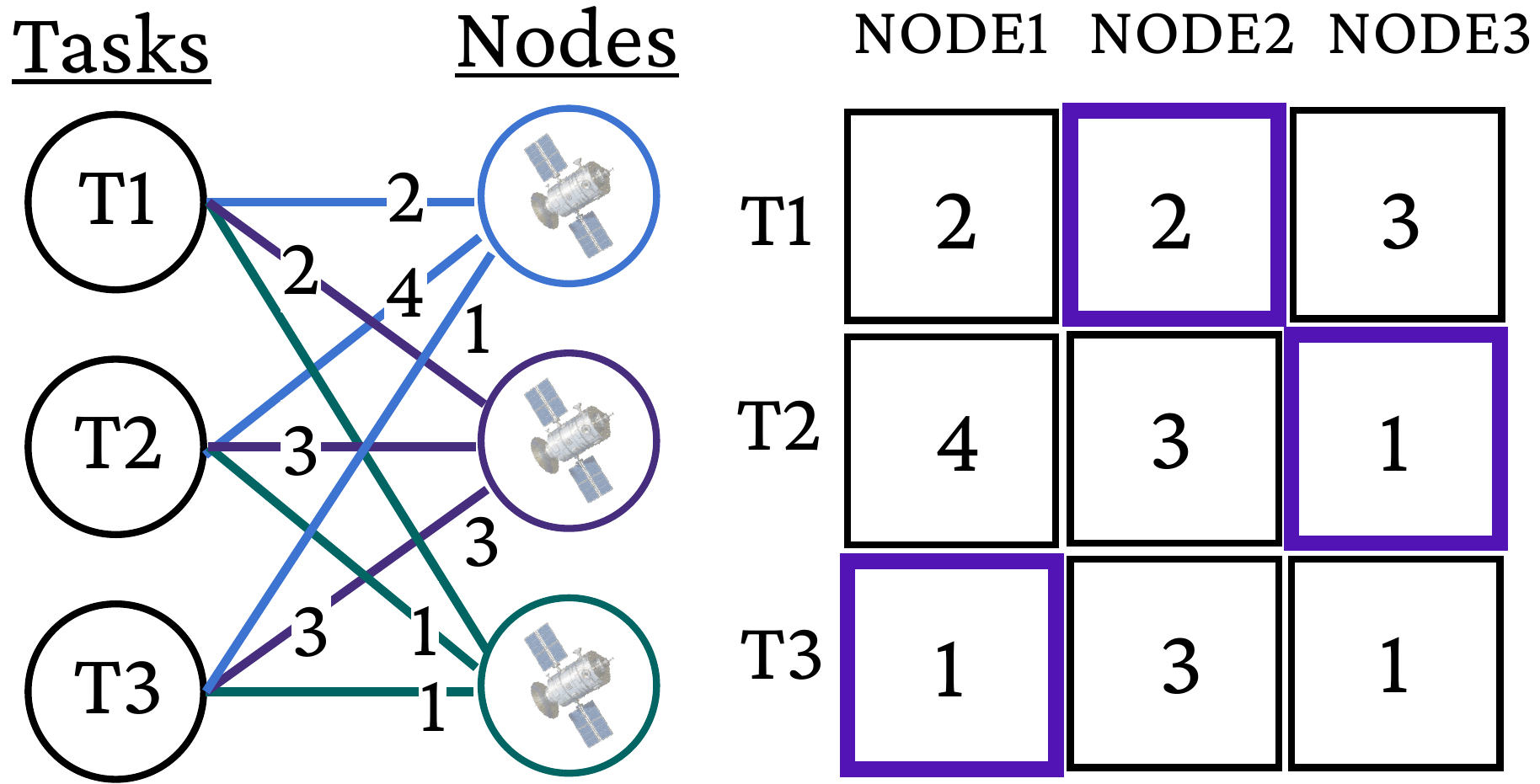}
    \caption{Task Processor Cost Adjacency Matrix Example}
    \label{fig:taskprocmodel}
\end{figure}

We formalize the orbital network structure and task scheduling optimization that underpin SpaceCoMP.

\subsection{Orbital Network Model}
\label{sec:orbital-model}

We assume a Walker Delta constellation shell comprising $N$ orbital planes with $M$ satellites per plane. Satellites are identified by tuples $(s, o)$ where $s \in \{1, \ldots, M\}$ indexes position within a plane and $o \in \{1, \ldots, N\}$ indexes the orbital plane itself. All satellites share the same altitude $h$ and inclination $i$, with planes uniformly distributed around the equator at $360^{\circ}/N$ separation. Within each plane, satellites are equidistant along the orbital path.

\subsubsection{Network Topology}

Each satellite maintains four inter-satellite links to its nearest neighbors: two within the same orbital plane (ahead and behind) and two in adjacent planes (left and right), forming a +Grid (2D-Torus) topology.

The \textbf{intra-plane distance} between adjacent satellites is constant:
\begin{equation}
D_m = (r_E + h)\sqrt{2\left[1 - \cos\frac{2\pi}{M}\right]}
\label{eq:intra-plane}
\end{equation}
where $r_E = 6{,}371$~km is Earth's radius. The \textbf{inter-plane distance} varies over the orbital period $T$ as planes converge near the poles and diverge at the equator:
\begin{equation}
\begin{aligned}
D_n(t) &= (r_E + h)\, \sqrt{2\!\left[1 - \cos\!\left[\frac{2\pi}{N}\right]\right]} \\
       &\qquad\times \sqrt{ \cos^2\!\left[2\pi\frac{t}{T}\right]  +\; \cos^2(i)\, \sin^2\!\left[2\pi\frac{t}{T}\right] }
\label{eq:inter-plane}
\end{aligned}
\end{equation}
where $t \in [0, T)$ is the time since the satellite crossed the equator (ascending), and the orbital period is:
\begin{equation}
T = 2\pi\sqrt{\frac{(r_E + h)^3}{\mu}}
\label{eq:period}
\end{equation}
with $\mu \approx 3.986 \times 10^{14}$~m$^3$/s$^2$ (Earth's gravitational parameter).

\subsubsection{AOI$\rightarrow$Satellite Mapping}

An Area of Interest is specified as a geographic bounding box: upper-left and lower-right (latitude, longitude) pairs. We map satellite positions to ground coordinates using the Earth-Centered, Earth-Fixed (ECEF) system and the SGP4 propagation model~\cite{vallado2006revisiting}.

A satellite may be in one of three states relative to an AOI at time $t$: \emph{within AOI ascending}, \emph{within AOI descending}, or \emph{outside AOI}. To ensure stable ISL connectivity, SpaceCoMP selects only ascending \emph{or} only descending satellites for any computation, never both, as their relative velocity ($>$15~km/s) prevents reliable links.

\subsection{Task Scheduling Model}
\label{sec:scheduling}

The core optimization goal is minimizing data movement: placing map tasks close to collectors, and the reducer close to mappers.

\subsubsection{Problem Formulation}

Given an AOI, let $\mathcal{T} = \{t_1, \ldots, t_k\}$ denote collector nodes (data sources) and $\mathcal{P} = \{p_1, \ldots, p_k\}$ denote candidate mapper nodes (processors). We seek an assignment $\sigma: \mathcal{T} \rightarrow \mathcal{P}$ minimizing total placement cost:
\begin{equation}
\min_{\sigma} \sum_{t \in \mathcal{T}} C(t, \sigma(t))
\label{eq:assignment}
\end{equation}
subject to each mapper processing exactly one task (bijective assignment).

This is the \emph{linear sum assignment problem}, solvable in $O(k^3)$ time via the Hungarian algorithm~\cite{munkres1957}. The problem can equivalently be viewed as finding a minimum-weight perfect matching in a bipartite graph where edge weights represent placement costs.

\subsubsection{Cost Function}

The cost of assigning task $t$ to processor $p$ comprises processing time and data transfer time:
\begin{equation}
C(t, p) = m_p \cdot K + h_{t,p} \cdot H + S(d_{t \rightarrow p}, V)
\label{eq:cost}
\end{equation}
where:
\begin{itemize}
    \item $m_p$ is the map processing time factor
    \item $K$ is a normalization constant (processing cost per unit)
    \item $h_{t,p}$ is the hop count from $t$ to $p$
    \item $H$ is the per-hop overhead (queuing, forwarding)
    \item $S(d, V)$ is the transmission time for $V$ bytes over distance $d$
\end{itemize}

The transmission time accounts for propagation delay and channel capacity:
\begin{equation}
S(d, V) = \frac{d}{\textcolor{violet}{c}} + \frac{V}{B \log_2(1 + \text{SNR}(d))}
\label{eq:transmission}
\end{equation}
where \textcolor{violet}{$c$} is the speed of light, $B$ is channel bandwidth, and SNR decreases with distance due to free-space path loss:
\begin{equation}
\text{SNR}(d) = \frac{P \cdot G_t \cdot G_r}{N \cdot \text{FSPL}(d)}, \quad \text{FSPL}(d) = \left(\frac{4\pi d}{\lambda}\right)^2
\label{eq:snr}
\end{equation}
with transmit power $P$, antenna gains $G_t, G_r$, wavelength $\lambda$, and noise power $N = k_B \cdot N_T \cdot B$ (Boltzmann constant $k_B$, noise temperature $N_T$).

The distance $d_{t \rightarrow p}$ is the sum of individual link distances along the Manhattan routing path, computed using Equations~\ref{eq:intra-plane} and~\ref{eq:inter-plane} for within-plane and cross-plane hops respectively.

\subsubsection{Reduce Placement}
\label{sec:reduce-placement}

After map tasks complete, their outputs must be aggregated at a reducer. We consider two placement strategies:

\textbf{LOS placement}: The reducer is the Line-of-Sight node coordinating the job. This minimizes the final hop to ground but may incur high aggregate transfer cost from distant mappers.

\textbf{Center-of-AOI placement}: The reducer is placed at the geometric center of the mapper distribution, minimizing total transfer distance from mappers. The reduced output then travels to the LOS node.

Overall, Center-of-AOI placement is advantageous when the reduce function significantly compresses data (high \emph{reduction factor} $F_R$), as the long reducer$\rightarrow$LOS transfer carries less volume.

%% file: simulation.tex
\section{Evaluation}
\label{sec:evaluation}

We evaluate SpaceCoMP through simulations modeling Walker Delta constellations with realistic ISL parameters. Our experiments address three questions:

\begin{enumerate}
    \item Does distance-aware routing reduce transmission cost without increasing hop count?
    \item Does bipartite matching outperform simpler task allocation strategies?
    \item Does center-of-AOI reduce placement improve end-to-end cost?
\end{enumerate}

\subsection{Simulation Setup}
\label{sec:eval-setup}

Table~\ref{tab:parameters} summarizes simulation parameters. We model constellations ranging from 1,000 to 10,000 satellites across 50--100 orbital planes at 87° inclination (typical for polar Earth observation missions). The AOI is defined as the satellite footprint over a large geographic region (United States) to ensure sufficient collector and mapper counts. For each configuration, we report means over 20 independent runs with randomized LOS node selection from cities with population exceeding 1 million.

We select 1/5 of AOI nodes as mappers and a non-overlapping 1/5 as collectors. All mappers and collectors are within the AOI, and we ensure only ascending (or only descending) satellites are selected per job.

\begin{table}[t]
\centering
\caption{Simulation Parameters}
\label{tab:parameters}
\renewcommand{\arraystretch}{1.3}
\setlength{\tabcolsep}{6pt}
\begin{tabular}{@{}clp{4.2cm}@{}}
\toprule
\textbf{Symbol} & \textbf{Value} & \textbf{Description} \\
\midrule
$i$ & 87° & Orbital inclination \\
$h$ & 530 km & Altitude \\
$B$ & 10 GHz & ISL channel bandwidth \\
$P$ & 5 W & Transmit power \\
$G_t, G_r$ & 62.5 dBi & Transmit/receive antenna gain \\
$N_T$ & 300 K & Noise temperature \\
$\lambda$ & 1550 nm & Wavelength \\
$V$ & 10 GB & Data volume per collect task \\
$F_R$ & 5 & Reduce compression factor \\
$F_M$ & 1 & Map compression factor \\
$m_p, r_p$ & 1 & Map/reduce processing time factors \\
$t_h$ & 3 & Processing overhead per hop \\
\bottomrule
\end{tabular}
\end{table}

\subsection{Distance-Optimized Routing}
\label{sec:routing-eval}

Standard +Grid routing selects the shortest Manhattan path by hop count, ignoring physical distance. Since inter-plane link distances vary by up to 40\% over an orbital period (Equation~\ref{eq:inter-plane}), we propose a \emph{distance-minimizing, hop-preserving} routing algorithm.

\subsubsection{Algorithm}

Each routing decision uses only local information: a node knows its four neighbors, constellation parameters ($M$, $N$, $i$, $h$), and current time. The algorithm proceeds:

\begin{enumerate}[i]
    \item Compute shortest horizontal (cross-plane) and vertical (within-plane) directions to destination.
    \item Compute inter-plane distances in forward, backward, and current positions.
    \item If at polar crossover (both directions larger than current), route horizontally to exit.
    \item If forward inter-plane distance is smaller than current, route horizontally.
    \item Otherwise, route vertically to defer cross-plane hops until links shorten.
\end{enumerate}

\subsubsection{Results}

\begin{figure}[t]
    \centering
\includegraphics[width=0.92\columnwidth]{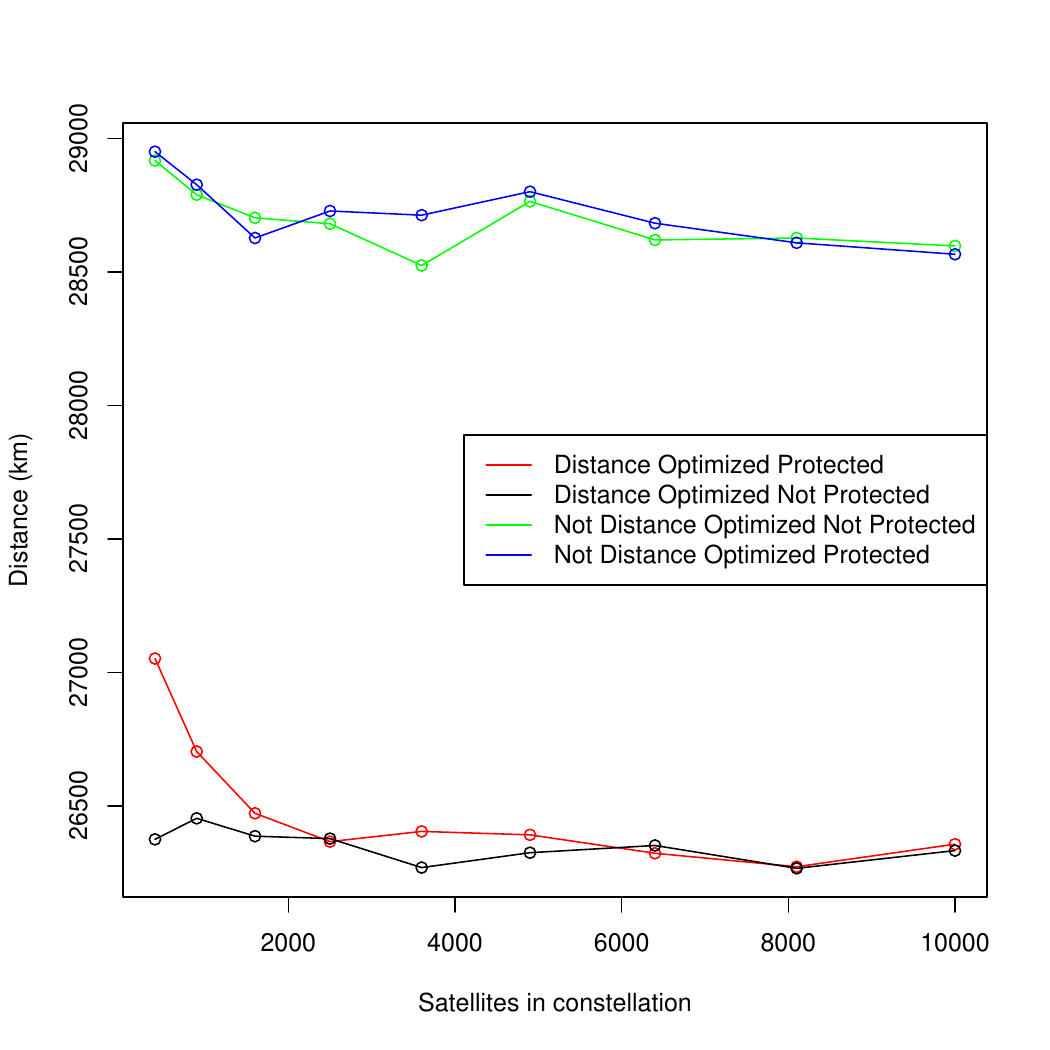}
    \caption{Routing distance vs.\ constellation size. Distance optimization reduces path length by 8--10\% at 53° inclination, up to 21\% at 87°.}
    \label{fig:routing-distance}
\end{figure}

\begin{figure}[t]
    \centering    \includegraphics[width=0.92\columnwidth]{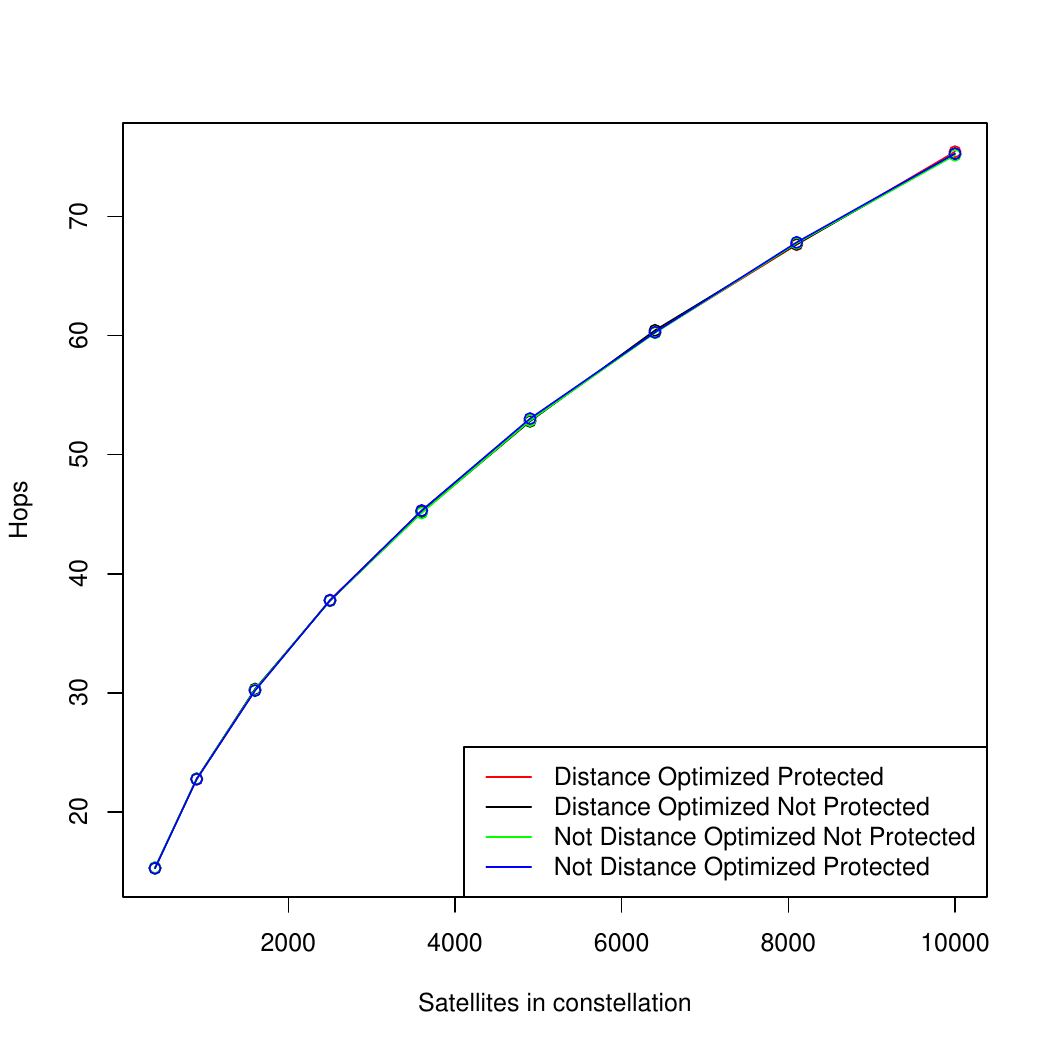}
    \caption{Hop count comparison. Distance optimization preserves hop count while reducing physical distance.}
    \label{fig:routing-hops}
\end{figure}

Figures~\ref{fig:routing-distance} and~\ref{fig:routing-hops} show that distance optimization reduces average path length by 8--10\% at 53° inclination and up to 21\% at 87° inclination, without increasing hop count. Higher inclination amplifies the benefit because inter-plane distance variation is more pronounced near poles.

\subsection{Map Task Allocation}
\label{sec:map-eval}

We compare three allocation strategies:

\begin{itemize}
    \item \textbf{Random}: Each task assigned to a uniformly random available mapper.
    \item \textbf{Eager}: Tasks processed sequentially; each assigned to lowest-cost available mapper.
    \item \textbf{Bipartite}: Optimal assignment via Hungarian algorithm on full cost matrix.
\end{itemize}

\begin{figure}[t]
    \centering
\includegraphics[width=0.92\columnwidth]{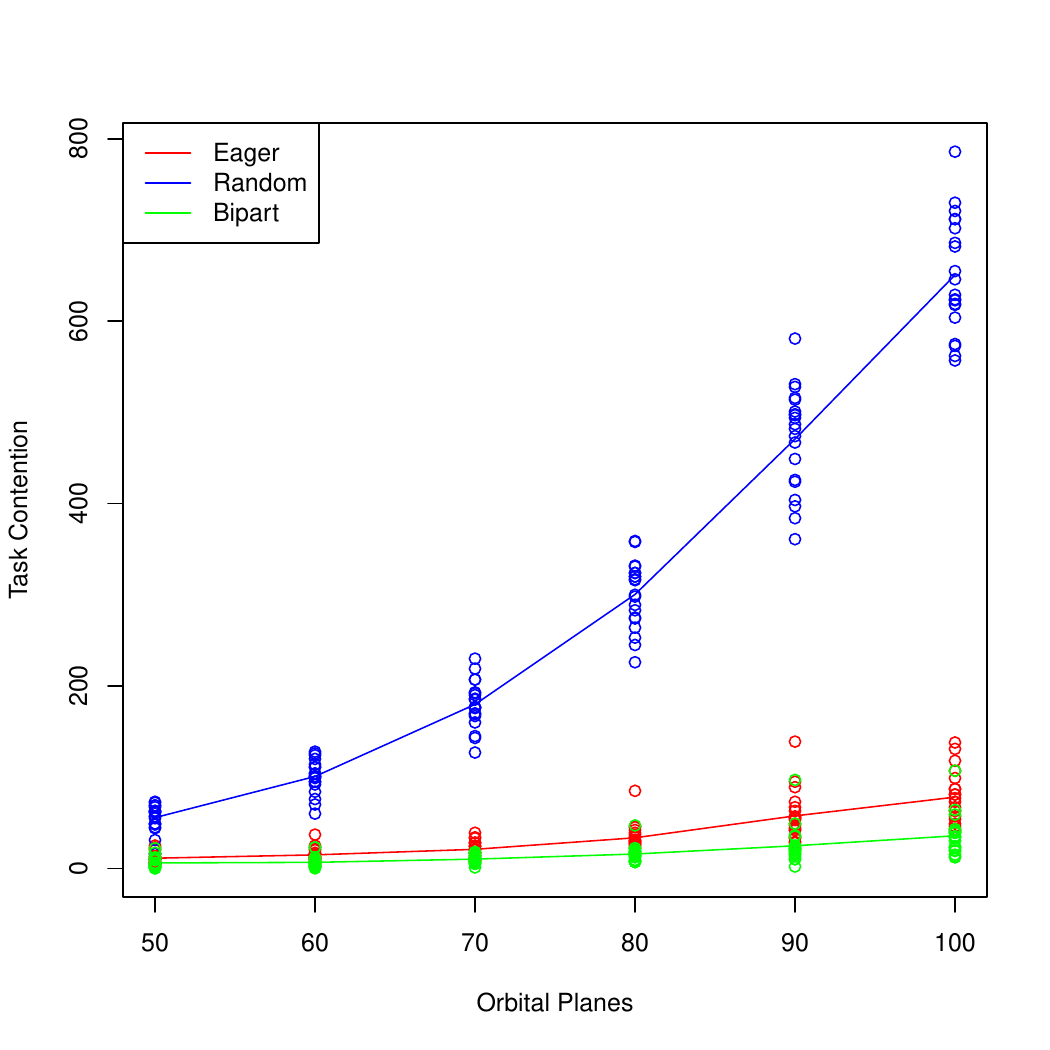}
    \caption{Map allocation improvement. Bipartite matching achieves 61--79\% improvement over random, 18--28\% over eager.}
    \label{fig:map-allocation}
\end{figure}

\subsubsection{Results}

Figure~\ref{fig:map-allocation} shows that bipartite matching achieves \textbf{61--79\% improvement} over random allocation and \textbf{18--28\% improvement} over eager allocation. The eager strategy degrades as task count increases because early greedy choices preclude globally optimal assignments.

\subsection{End-to-End Cost}
\label{sec:e2e-eval}

We evaluate complete SpaceCoMP jobs combining all optimizations.

\subsubsection{Map Phase}

\begin{figure}[t]
    \centering
\includegraphics[width=0.92\columnwidth]{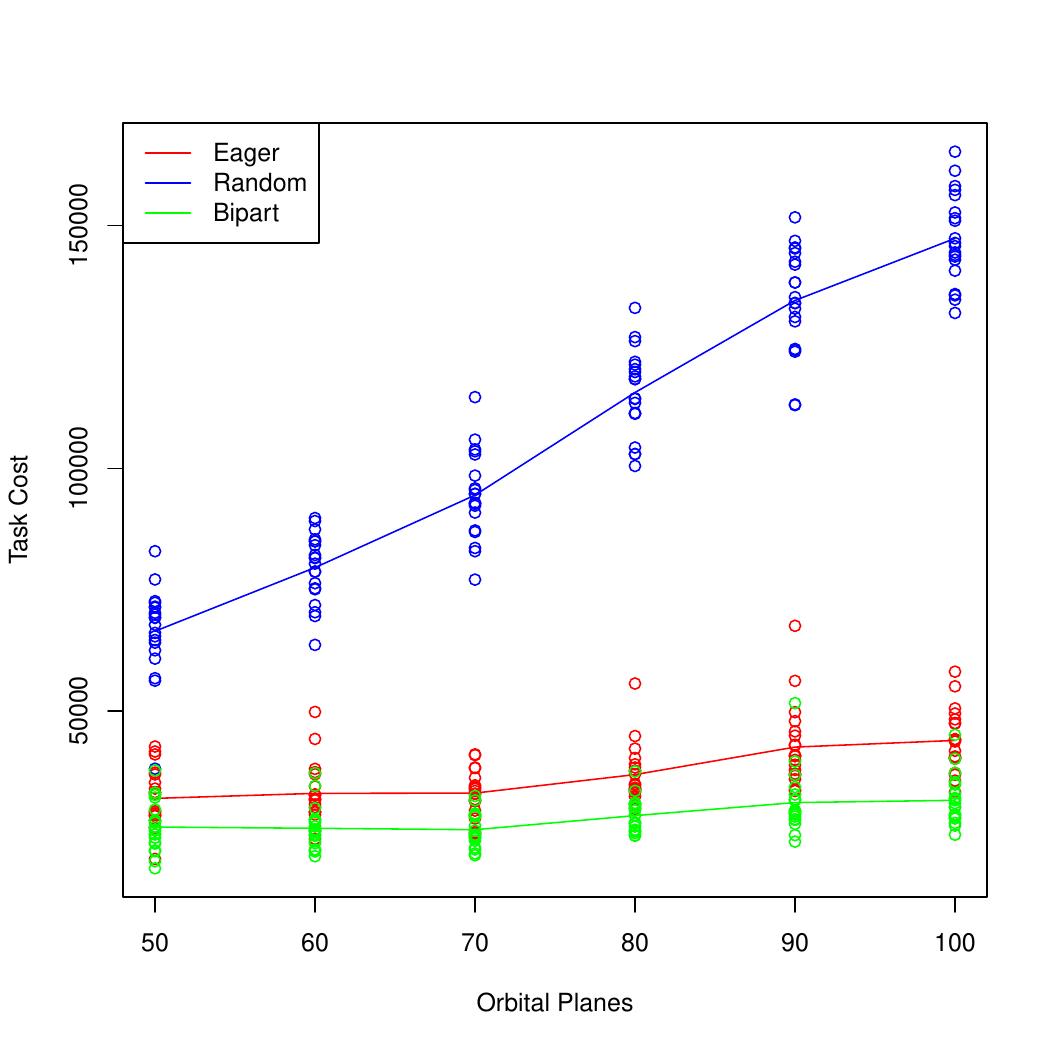}
    \caption{End-to-end map cost. Bipartite allocation consistently outperforms eager and random baselines.}
    \label{fig:e2e-map}
\end{figure}

Figure~\ref{fig:e2e-map} shows end-to-end map cost as orbital plane count increases. Bipartite allocation consistently outperforms baselines across all constellation sizes.

\subsubsection{Reduce Phase}

\begin{figure}[t]
    \centering
\includegraphics[width=0.92\columnwidth]{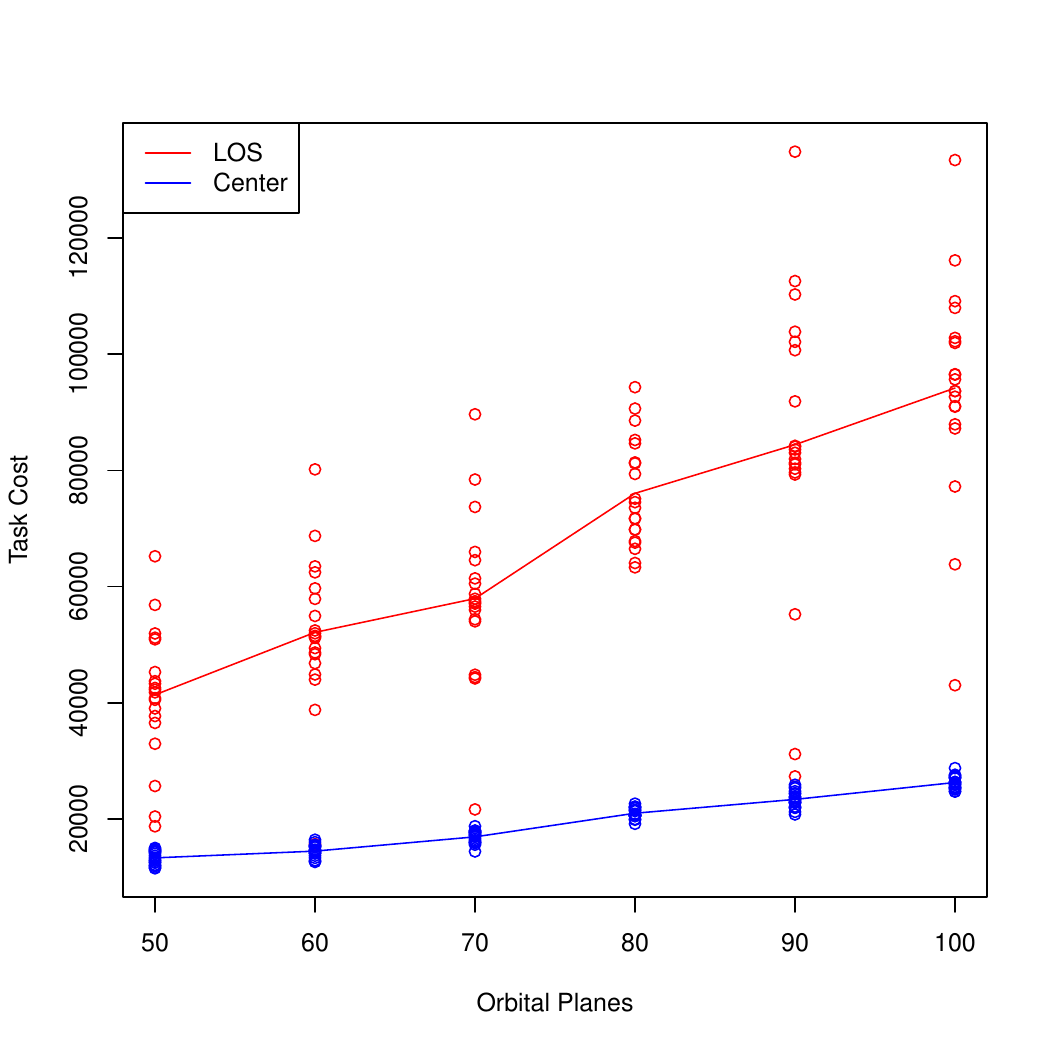}
    \caption{End-to-end reduce cost. Center-of-AOI placement achieves 67--72\% lower cost than LOS placement.}
    \label{fig:e2e-reduce}
\end{figure}

Figure~\ref{fig:e2e-reduce} compares LOS and center-of-AOI reduce placement. Center placement reduces cost by \textbf{67--72\%} at reduction factor $F_R = 5$.

\subsubsection{Impact of Reduction Factor}

\begin{figure}[t]
    \centering
    \includegraphics[width=0.92\columnwidth]{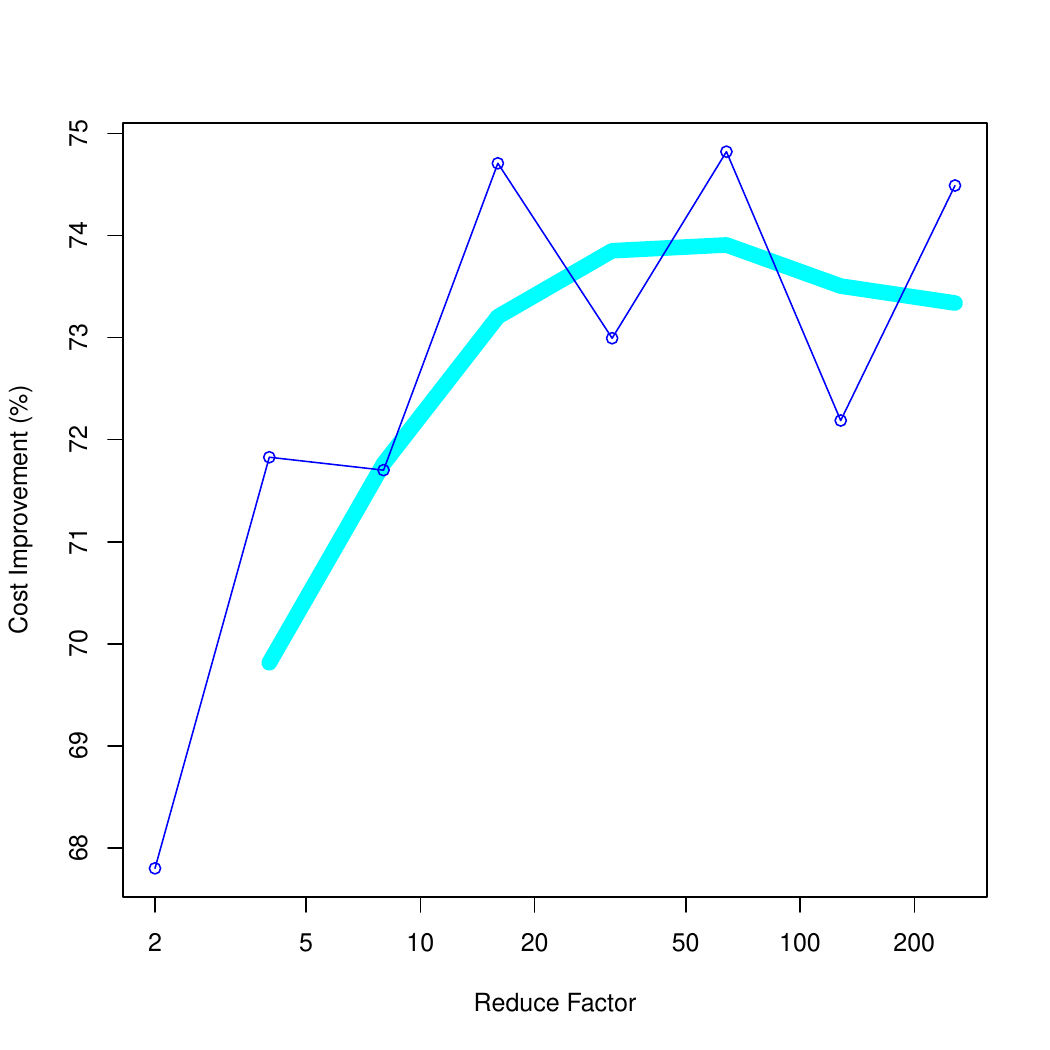}
    \caption{Reduce cost improvement vs.\ reduction factor. Benefits increase with compression, saturating around $F_R = 50$.}
    \label{fig:reduce-factor}
\end{figure}

Figure~\ref{fig:reduce-factor} shows how benefits vary with reduction factor $F_R$. At $F_R = 1$, center placement offers modest benefit; as $F_R$ increases, benefits grow and saturate when mapper-to-reducer transfers dominate.

\subsubsection{Contention Analysis}

\begin{figure}
    \centering
    \includegraphics[width=0.92\linewidth]{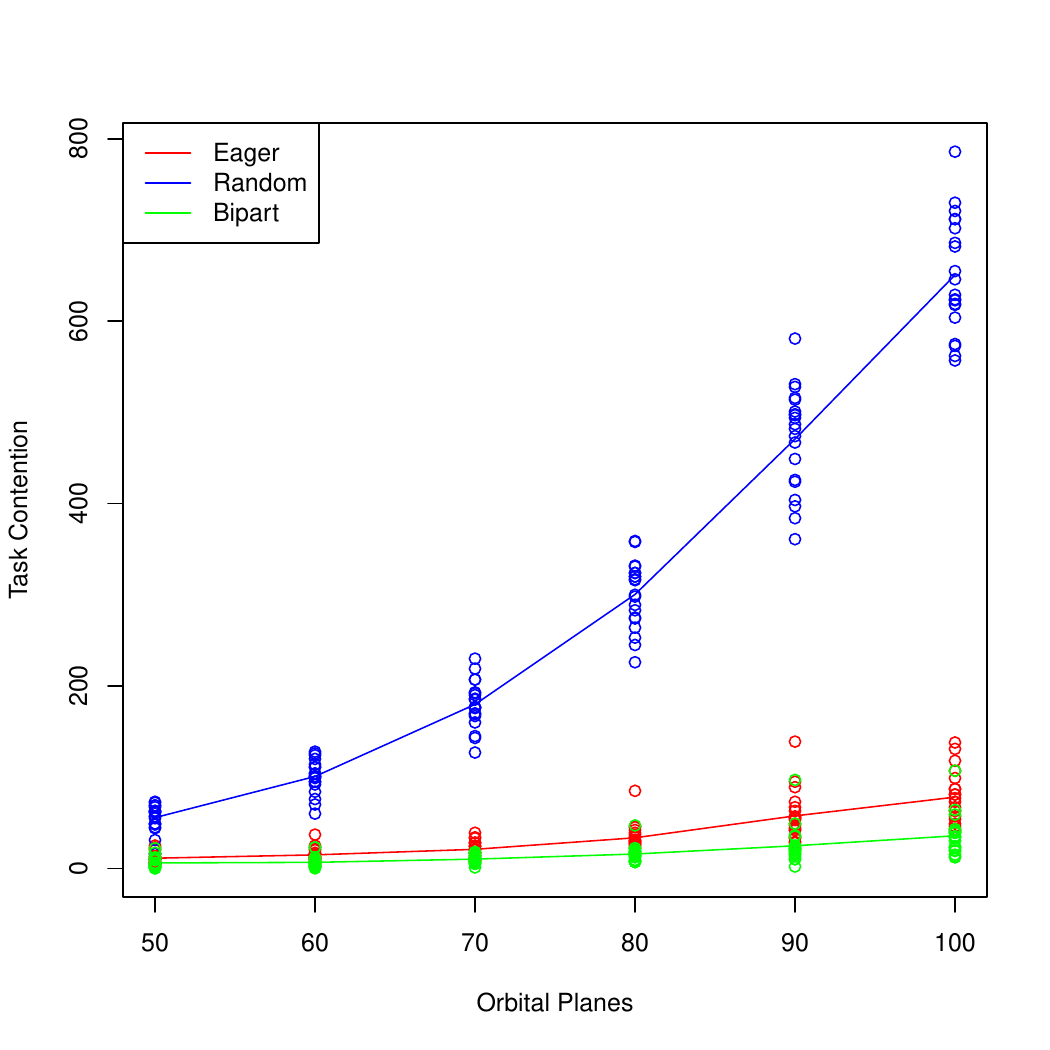}
    \caption{Map phase contention. Bipartite allocation distributes load more evenly.}
    \label{fig:e2emapcont}
\end{figure}

\begin{figure}
    \centering
    \includegraphics[width=0.92\linewidth]{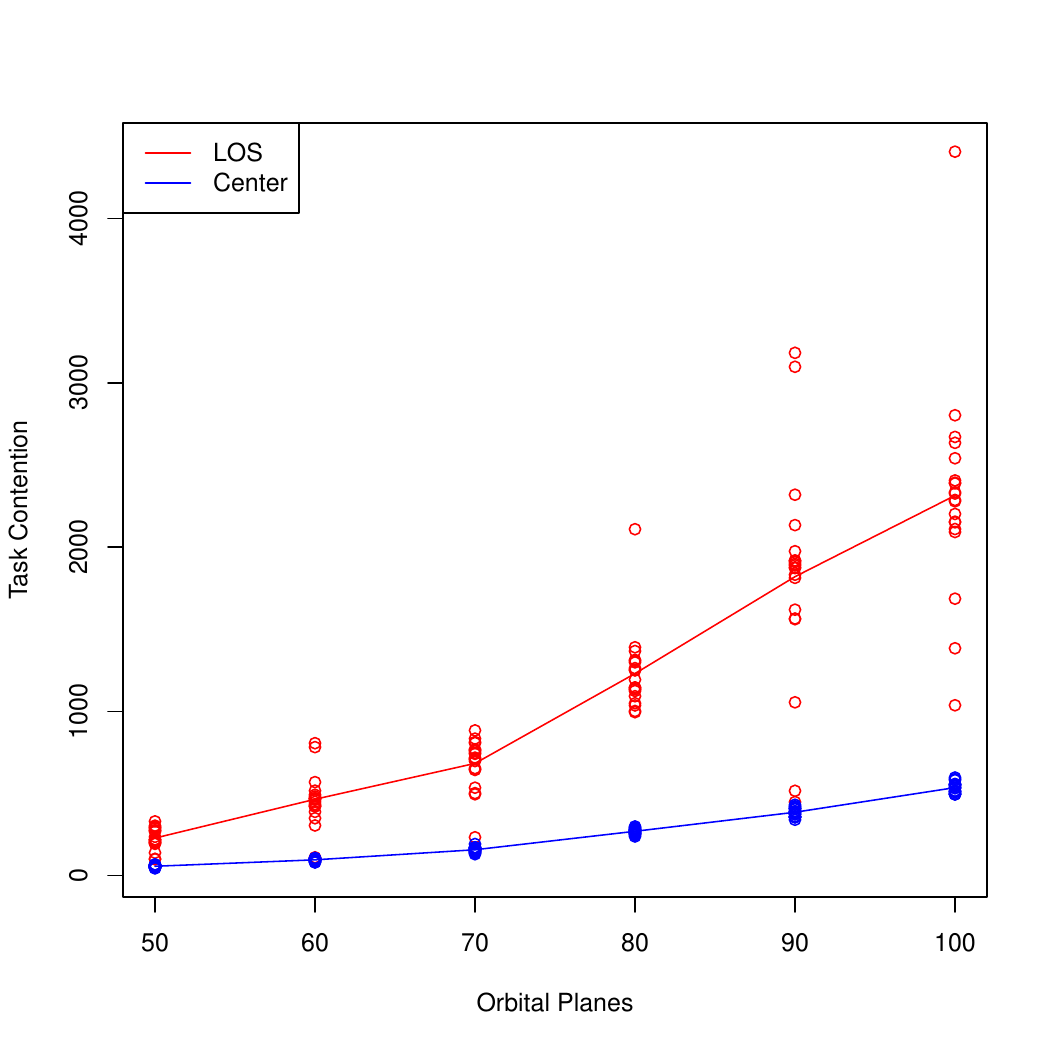}
    \caption{Reduce phase contention. Center placement reduces hot-spot formation.}
    \label{fig:e2ereducecont}
\end{figure}

Figures~\ref{fig:e2emapcont} and~\ref{fig:e2ereducecont} show node visit counts as a proxy for congestion. Bipartite allocation and center placement both reduce contention by distributing traffic more evenly.

\subsection{Summary of Findings}

SpaceCoMP's optimizations yield compounding benefits:
\begin{itemize}
    \item \textbf{Routing}: 8--21\% distance reduction with zero hop overhead
    \item \textbf{Map allocation}: 61--79\% over random, 18--28\% over greedy
    \item \textbf{Reduce placement}: 67--72\% cost reduction vs.\ LOS placement
\end{itemize}

%% file: conclusion.tex
\section{Discussion and Conclusion}
In this paper we presented a method similar to  tools like MapReduce, in order to simplify the development and deployment of distributed space computing
computations. We did so by offering services that take advantage of
the unique properties of LEO ISL networks, such as orbital
dynamics. Thus, applications do not have to re-implement these
features. 

Furthermore, we discussed the architecture and baseline protocol,
as well as ran simulations to validate the model. Additionally, we
contributed a novel routing protocol, a map task allocation
scheduler, and a reduce placement strategy.

We expect this SpaceCoMP model to provide value for data-intensive
real-time analysis workloads, such as computer vision detection and classification tasks that can be triggered by ground MCP clients via LLM prompts. Those prompts include satellite data and intelligence in model context for dynamic inference combined with other agent data. Hence, the speed at which results can be delivered is of paramount importance, and horizontally scaling over a large number of nodes is an approach that has worked well in data-centers on the ground. 

Another interesting use case for satellite earth observation is multi-image super resolution (MISR),  whereby many satellites take lower resolution images that are then combined (reduced) into a single higher-resolution image. Clearly, doing the reduction work in orbit would save bandwidth capacity on the downlink.
Furthermore, using the approach of this model allows developers to focus on writing sensor collection code, map processing, and reduce aggregation code without the need to re-implement  distributed systems infrastructure such as fail-over, scaling and scheduling.

As mappers and reducers coalesce sensor data with technologies such as MISR~\cite{farsiu2004fast}, improved synchronization results in better resolution. Classical clock synchronization protocols remain vulnerable to spoofing and delay attacks, as timing information can be copied or manipulated in transit. Indeed, successive versions of PTP have required security patches as new attacks are discovered. Quantum time synchronization offers a promising solution, and future extension to our protocol, because of the non-cloning nature of quantum states and their inherent resistance to eavesdropping. In addition, quantum  mechanisms allow for timing precisions of the order of picoseconds, vastly improving the computing efficiency of the satellite network. Equally important, these quantum mechanisms can be deployed without requiring a master clock outside the LEO constellation.

We also note that recent developments to upgrade satellite hardware
to run AI workloads more efficiently, such as neuromorphic computing and
vector processing, will help develop more sophisticated map (and reduce)
tasks~\footnote{https://www.nasa.gov/game-changing-development-projects/high-performance-spaceflight-computing-hpsc/}. This part of the recent trend of using space networks and distributed systems running through them in order  to disaggregate and simplify space flight system components. This in turn leads to more cost-effective launches and operations, as seen by the improvements achieved in recent years~\cite{crum2022}.

Future work will include implementing the proposed model on the NASA cFS platform\footnote{https://etd.gsfc.nasa.gov/capabilities/core-flight-system/} using the ColonyOS meta operating system~\cite{kristiansson2024}. In our simulations we assumed static costs for running tasks on satellites, whereas real deployments there would be multiple competing task schedulers using the same nodes in order to run multiple tasks. As a consequence we envision embedding more dynamic information in the Task Processor Cost Adjacency Matrix exhibited above such as current load and processing queue size. We could also more explicitly embed cost of contention for each allocation to avoid bottleneck links.

\section*{Acknowledgments}
This work was supported and funded by the Swedish National Space Agency (SNSA) as part of the LeoDOS project (https://www.leodos.org) under contract DNR 2025-00306.